# COVID-19 Contact-Tracing Mobile Apps:
## Evaluation And Assessment For Decision Makers

# Table Of Contents

# The Case For Action

# Evaluating Contact-Tracing Mobile Apps:
## The Questions We Are Asking

# Conclusion

# References

# The Case
# For Action

More than 150,000 deaths are now attributed to the global COVID-19 pandemic. Many thousands more lives are expected to be lost before we have brought the disease under control and are capable of managing future spikes in the number of cases. In an effort to both slow and stop the disease, communities across the world have halted everyday life, requesting or requiring their residents to close non-essential businesses, stop going to school, and stay home. Digital initiatives hope to support safe and well-considered approaches to the reopening of our societies while simultaneously reducing the human loss of life by giving frontline officials modern tools with which to control this pandemic.

One particular set of modern digital tools aims to upgrade contact-tracing capacity, typically a lengthy and laborious process. In addition to increasing the speed with which contact-tracers can reach those who have been exposed to the disease, these tools can increase the accuracy of contact tracing. However, many first-generation digital contact-tracing tools have paved the way for a post-pandemic surveillance state and the mistreatment of private, personal information. Privacy must remain at the forefront of the global response, lest short-term pandemic interventions enable long-term surveillance and abuse. The design and development of the next generation of contact-tracing tools offers an opportunity to sharply pivot to solutions using privacy-first principles and collaborative, open-source designs. These tools present an opportunity to save lives by flattening the curve of the pandemic and to provide economic relief without allowing privacy infringements now or in the future.

## The COVID-19 Pandemic Challenge

COVID-19 virus transmission occurs for several days before a person shows any symptoms. During this time, a person going about their daily life may interact with, and possibly pass the infection to, as many as a thousand people. Without knowing they are infected, an individual who has only mild symptoms or is asymptomatic may continue to interact with others, further spreading the virus. This creates an exponential rise in infections.

Stopping the spread of COVID-19 with pharmaceutical treatments and vaccines remains at least 6–18 months away from widespread availability. Therefore, public health countermeasures, such as social distancing, offer the only possibility of stopping virus proliferation in the near future. When applied broadly, such measures disrupt every aspect of society and risk economic collapse. Already, unemployment rates have skyrocketed, tenants are struggling to pay rent, and critical supply chains, including the food supply chain, have been interrupted. The longer strict social distancing measures remain in place, the more severe the consequences for economies and societies will be. However, if social distancing measures are lifted too quickly, the virus will spread once again, claiming many additional lives.

# The Contact-Tracing App As A Solution

The contact-tracing process evaluates the recent location history and social connections of those who become infected and notifies the people they have interacted with of their exposure to the virus. In this way, contact-tracing methods allow targeted measures (e.g., quarantining, virus testing) to be applied only to exposed individuals. Traditionally, public health officials perform contact tracing manually, by interviewing patients diagnosed with a disease about their activity over the past days or weeks. Then, officials reach out to people who crossed paths with the patient during the time the patient was contagious and recommend targeted interventions to prevent further spread of the disease.

Widespread, rapid transmission of a virus by respiratory droplets, as in the case of COVID-19, challenges the practicality of the traditional contact-tracing process. Manual tracing is resource intensive, is time consuming, and will, at best, be limited to contacts within the social circles of the infected—and thus cannot trace strangers effectively. Furthermore, the patient being interviewed is often extremely ill and at risk for memory errors during the interview. Digital contact-tracing tools may help mitigate these challenges.

Today, almost half of the world's population carries a device, such as a smartphone, capable of GPS tracking and Bluetooth communication with nearby devices. Each device is able to create a location trail—a timestamped log of the locations of an individual, as well as a list of anonymous ID tokens that are collected when the device user crosses near another device. By comparing the device users' location trails or the anonymous ID tokens they have collected with those from people who have COVID-19, one can identify others who have been near the person who is infected; this facilitates contact tracing in a more accurate and timely manner than the traditional manual approach. Several pilot programs, particularly in China and South Korea, have demonstrated the technical feasibility of contact-tracing applications as tools to help contain the COVID-19 outbreak within a large population. However, these programs also highlight the very real risks that exist with the use of such technologies.

A location trail and list of nearby device IDs contains highly sensitive, private information about a person: everything from where they live and work and which businesses they support, to which friends and family members they visit. Location data can be used to identify people who are infected and might then be targeted by their community. For example, data sent out by the South Korean government to inform residents about the movements of persons recently diagnosed with COVID-19 sparked speculations about the individuals' personal lives, from rumors of plastic surgery to infidelity and prostitution. More frightening still, enabling access to a person's location data by a third party, particularly a government, opens a path to potentially unrestrained state surveillance. In China, users suspect that an app developed to help citizens identify symptoms and their risk of carrying a pathogen was used to spy on them and share personal data with the police. Care must be taken in the design of such apps.

A number of groups, from governments to non-profits, have quickly acted to innovate the contact-tracing process: they are designing, building, and launching contact-tracing apps in response to the COVID-19 crisis. A diverse range of approaches exist, creating challenging choices for officials looking to implement contact-tracing technology in their community and raising concerns about these choices among citizens asked to participate in contact tracing. We are frequently asked how to evaluate and differentiate between the options for contact-tracing applications. Here, we share the questions we ask about app features and plans when reviewing the many contact-tracing apps appearing on the global stage.

"

## [Some of my patients] were more afraid of being blamed than dying of the virus"

— <u>Lee Su-young</u>, Psychiatrist at Myongji Hospital, South Korea

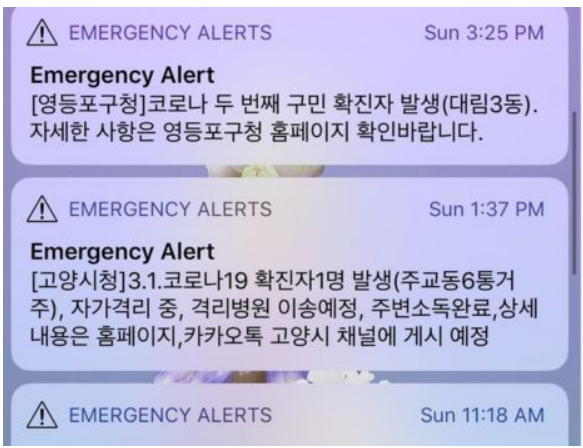

*Examples of the messages that alert South Koreans to new cases*

# Contact-Tracing Apps Around The Globe

# North America
## USA
COVID Safe Paths

CoEpi

Covid Watch

# South America
## Peru
PeruEnTusManos ("Peru in Your Hands")

# Asia
## Singapore
TraceTogether

## South Korea
Corona 100m (Co100)

## China
Alipay Health Code, WeChat, QQ

## India
Aarogya Setu

Tracy

# Europe
## UK
Covid Symptom Tracker

## France
StopCovid

## Poland
ProteGO

## Iceland
Rakning C-19

## Croatia
STOP Corona!

## Austria
Stopp Corona

# Middle East
## Israel
HaMagen

Track Virus

## Bahrain
BeAware Bahrain

# Africa
## South Africa
Covi-ID

Additional proposed apps beyond this list may exist. All apps listed are not necessarily available in Apple iOS store and Google Playstore at this time. As part of this paper, we have begun to seek answers to the questions that follow for each of these apps. These answers are being gathered into a table; the working draft can be found here:

https://docs.google.com/spreadsheets/tracing_apps_table

We invite readers to submit additions and suggestions to the table at info@pathcheck.org

# The Questions We Are Asking

## Is The Project Open Source?

An open-source approach lets programmers and other experts outside the app development team review the code for a project. These outside programmers can make improvements, copy the code, or use it to create something entirely new. Open source offers a layer of trustworthiness. Because the code is publicly available, it can be reviewed by experts around the world to confirm it works the way the development team says it should. There are, at times, valid reasons to not use an open-source approach, such as when a business is seeking to develop a proprietary technology.

During the COVID-19 crisis, we believe that open-source projects promote collaboration and foster community.

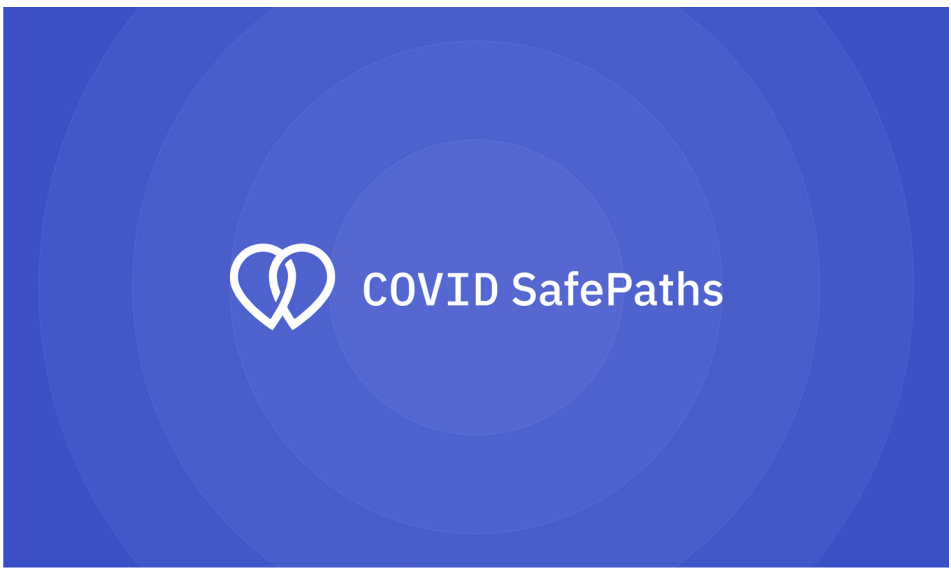

## Does the App Use GPS, Bluetooth, or Both?

Contact-tracing apps require the use of a data source to infer contact between two people: two of the most useful data sources are GPS location data and Bluetooth broadcasting. GPS-based apps create a "location trail" for each user by recording their time-stamped GPS location. If a person catches COVID-19, they can share their location trail with the responsible authority—the health worker, public health official, government official, or app creator. The authority then releases some or all of the location trail for other users to compare to. In some applications, the person who is infected might be able to directly share their location trail with other users.

Other apps rely on Bluetooth to determine who the person who is infected has crossed paths. Such apps create a unique identifier, a number or token, which the app broadcasts to nearby devices. The user's phone then records the identifiers of other phones it has been near. If a person becomes infected, their unique identifiers can be compared to those stored by other users to determine who the infected person has crossed paths with. In some cases, such as the Singapore TraceTogether app, the central authority stores user information and can determine the user's phone number and identity from an identifier. In others, such as COVID Watch and CoEpi, the identifiers provided by the person who declares themselves to be infected cannot be used by the central authority to determine the person's real world identity.

Both approaches offer distinct advantages and challenges:

### GPS-Based Approach

- Allows for estimation of exposure related to surface transmission of disease. Unlike Bluetooth, GPS-based systems can notify users if they were in a location shortly after a person infected with COVID-19, when the chance for exposure to the virus through commonly touched surfaces is high.

- Enables users to import historical data. Other applications on the users' phones, such as Google Maps, are already collecting the potential user's location histories before they install the contact-tracing app. When users import these historical data, the app can alert the user to potential exposures from their location history, even before they downloaded the app.

- Provides redacted, anonymized GPS data to help public health officials follow the spread of disease within a community.

- Is able to record the user's location history using a small amount of data, making scaling and implementation in regions with high data costs more likely.

### Bluetooth-Based Approach

- Uses signal strength, which is reduced by walls and other barriers, to estimate the distance between users. In some places, such as a large, multi-floor building, this estimate more accurately reflects the chance of exposure to disease than a GPS-based approach.

- Uses time-range-dependent, randomly generated numbers as IDs to ideally achieve relative anonymity.

- Requires the use of a compatible app by other users to record possible exposures. If an app is not widely adopted, the potential utility is limited.

- No potential to collect historical data from before the user downloaded the app.

In the near-future, some solutions, including COVID Safe Paths, will integrate both approaches, allowing the user to harness the advantages of each while mitigating some challenges.

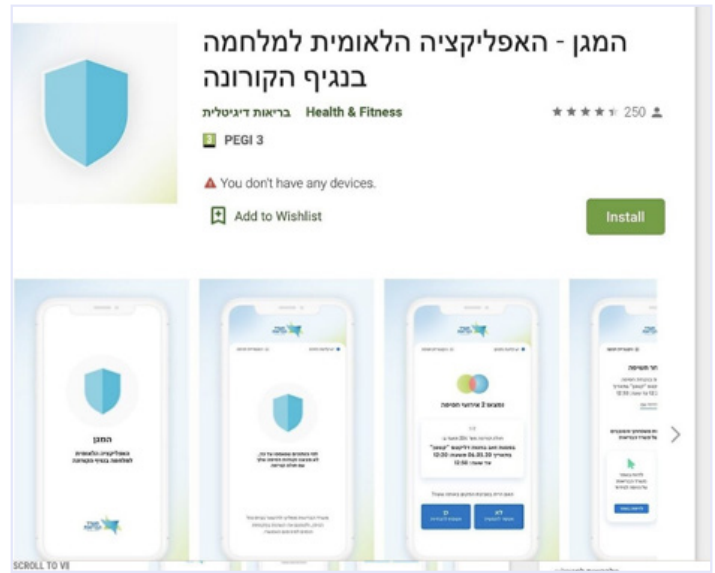

**GPS:** HaMagen (Israel)

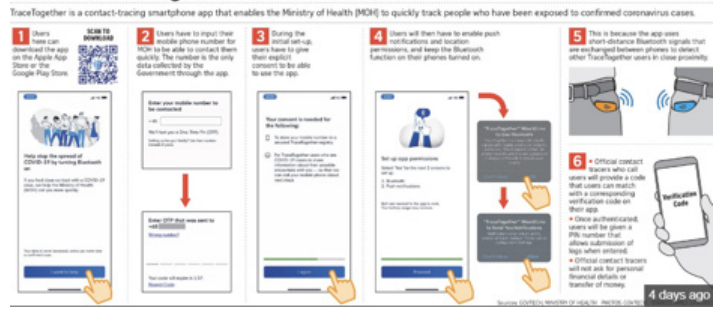

**Bluetooth:** Trace Together (Singapore)

**Fixed vs. Time-Variable Identifiers**
Some Bluetooth-based apps use a fixed identifier, meaning the unique number assigned to the device does not change and is permanently associated with the user. Time-variable identifiers change on a set time interval, such as an hour, so each user is associated with many different identifiers. The use of time-variable identifiers adds a layer of privacy protection by making it difficult for a third party to track a particular phone over time based upon a single identifier.

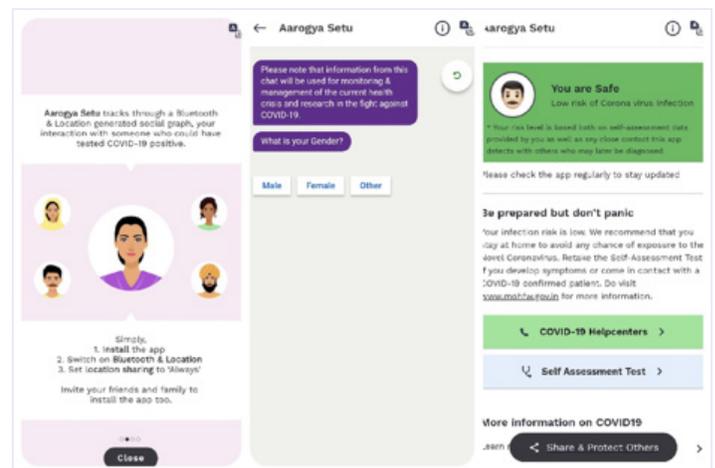

**Both GPS and Bluetooth:** Aarogya Setu (India)

## Are Location And Contact Data Stored And Processed In A Centralized Or Decentralized Manner?

In a centralized version of contact tracing, location and contact data are collected and consolidated centrally by a single authority, often a government entity. China utilized a centralized approach with its app. Other information about the user, such as mobile telecommunication service provider or payment data, may be collected and paired with the location data.

The central authority identifies people who are infected, determines their contacts, and requests specific actions by those who may have been exposed to the virus. Centralized systems create powerful tools for analysis and public health decision making. However, such systems also expose a person's data to a central authority, creating an opportunity to undermine the person's privacy.

In a decentralized approach, the healthy user's data never goes to a central server. Location data are stored and processed on the phone of the user. Only the location data of people confirmed to be infected need to be shared. Tools, such as redaction and blurring of the infected person's data, can be used to help preserve their privacy. An Israeli app, Track Virus, is an example of a decentralized approach, as is COVID Safe Paths. Decentralized systems typically offer greater privacy protection and are, therefore, more in line with privacy requirements and regulations such as GDPR. Some utility may be lost compared to centralized systems as collection and aggregation of large data sets from users can be used for beneficial public health research. However, as we consider the various approaches, the grave privacy risks associated with centralized systems far outweigh the limited additional benefits, leading us to highly value decentralized approaches.

*A previous version of this report incorrectly described Track Virus. Track Virus is a live platform with increasing use. It utilizes a decentralized approach to processing and storing data in order to preserve user privacy. More can be learned about Track Virus here:*

*https://www.track-virus.com/*

### A Push or a Pull?

When checking if a healthy user has been exposed to COVID-19, contact-tracing apps may either push the healthy user's data to the authority (centralized processing) or pull a list of locations and/or contact IDs of those who have been infected from the authority (decentralized processing). With a push, the healthy user's data is pushed (shared) off of the user's device and is compared by the authority to the data of people who have been infected. This exposes a large amount of data to the authority. In a pull model, an anonymized history of location data or identifiers from people who have been infected are pulled onto the healthy user's device so that the comparison can take place locally without compromising the privacy of healthy individuals.

## How Does the App Decide Which People Could Have Been Exposed To Others Infected With COVID-19?

At the base of every contact-tracing app lies an algorithm that determines whether the app user has been exposed to people who are infected and might have an increased chance of being infected themselves. The algorithm integrates many factors, such as the distance between the users, the length of time the users were in the same location, or the amount of time between the contact and the start of symptoms. Two apps with different algorithms will potentially give a different likelihood of exposure to the same user. Understanding the algorithm used is necessary for public health officials and healthcare providers to provide appropriate guidance to users who receive an exposure notification. Contact-tracing app developers must clearly communicate their algorithm with all stakeholders and failure to do so will be a significant red flag.

---

**Potential Exposures to COVID-19**

Given what is known to date about person-to-person transmission of COVID-19, contact-tracing apps can properly assess users' potential exposure to the virus if they take four important factors into consideration:

- The distance between the person who is infected and the user.

- The length of time the person who is infected and the user occupied the same space.

- How many days prior to becoming infected the person interacted with the user.

- Whether or not the user may have had contact with contaminated surfaces after interaction with the person who is infected.

---

## How Is Data Collected From People Who Experience Symptoms Of COVID-19?

A location history must be collected from a person who has been diagnosed with COVID-19 in order for contact tracing to occur. Several approaches are being piloted. In general, these approaches fall into two categories:

- An authority (public health official, healthcare provider, government official) collects the location history from the person who is infected and makes it available to users of the app.

- The patient self-reports symptoms and directly shares their data with other users of the app.

Use of an authority offers the advantage of confirmation that the person has COVID-19. The overlap of symptoms between COVID-19 and other common respiratory illnesses might cause someone to suspect they have COVID-19 when they actually have the flu or a common cold. Systems where people self-report themselves as infected pose the risk that people with symptoms, but without a confirmed diagnosis, share their location trail. Self-reporting approaches are also at risk from bad actors who may misreport their status as infected in order to create chaos and fear. However, self-reporting systems have the advantage of fuller consent of the infected person as the person definitively decides to share their location trail without influence from an authority figure. When evaluating contact-tracing solutions, we seek to understand how data will be collected from the person who is infected and how the solution will confirm that the person truly has COVID-19.

## Will The Data Be Used For Other Purposes Besides Contact Tracing?

Location data may potentially be repurposed to achieve additional objectives beyond contact tracing. We believe these data should be used only for response to an ongoing pandemic and that other uses should be strictly forbidden. Turning app data over to law enforcement or other non-health actors, such as commercial entities seeking to target ads to potential customers, threatens users' rights and privacy. Critically, this undermines public trust. Without trust, citizens will not adopt contact-tracing apps at a wide enough scale to effectively control the spread of the epidemic. Therefore, access to location-tracking data should be tightly limited to specific public health initiatives working on pandemic response.

## How Do Users Know The Data Is Not Used For Other Purposes?

Users should be able to confirm how their data is used. Promises by the app's developers to delete data are insufficient. Users should be able to check exactly what location data has been collected and stored and to confirm that their data is no longer there after the deadline for deletion (the disease's incubation period, 14 to 37 days for coronavirus). Apps must obtain users' unforced and informed consent for any disclosure of their data.

Recently, the A1 Teleom Austria Group shared aggregated user location data from an app not regularly used for public health purposes with the Austrian government's COVID-19 emergency management team for reasons that were not initially specified. Observers believe that A1's data was most likely being used to forecast disease spread or to monitor the population for large gatherings that might transmit the virus. However, the sharing of location data with government agencies for unspecified purposes attracted the criticism of privacy rights activists and created suspicions that weakened user trust, threatening long-term success.

## How Is The Privacy Of Healthy Users Protected?

An opportunity for misuse and privacy violations arises whenever a third party, a government, a corporation, or any other entity is able to access the data of healthy users. A decentralized approach prevents privacy compromise for healthy users because they are doing all the calculations on their own phones. Time-limited storage of location data also protects user privacy, such as only storing 28 days of data with deletion of everything beyond this point. All contact-tracing app development teams should clearly articulate how they protect the privacy of all users – whether healthy or infected. As an example, a preliminary draft of the privacy principles of the COVID Safe Paths team can be accessed in COVID-19 Contact Tracing Privacy Principles. This overview of model privacy practices explains how the application embraces principles such as Privacy by Design, the Fair Information Practice Principles (FIPPs), and Legal Protection by Design.

## How Is The Privacy Of Users Who Have Been Infected With COVID-19 Protected?

Historical location data and nearby device IDs must be collected from a person who is infected to enable contact tracing. However, both the collection and release of that information have broad implications for the privacy rights of the individual. As the most vulnerable stakeholder, several efforts must be undertaken to protect, to the highest degree possible, the privacy of the person who is infected.

App development teams may design for privacy by utilizing a variety of approaches:

- Limiting the amount of data published publicly.

- Providing tools that allow the person who has been diagnosed and their healthcare providers to redact any sensitive locations, such as a home or workplace.

- End-to-end encryption of location data before sensitive locations are redacted.

- Eliminating the risk of third-party access to information by enabling voluntary self-reporting by the person who is infected.

- Supporting strict regulation around access to and usage of the data by any entity that collects it, particularly governments.

- Obtaining targeted, affirmative, informed consent for each use of the person's data.

- Providing users with the ability to see how their data is being used and revoke consent for usage of their information.

- Providing users with the ability to correct incorrect information.

- Notifying individuals about what data is collected, how long it is stored, and who will have access to it during each stage of use.

- Enabling people to obtain access to information about potential exposures to COVID-19 without requiring that they consent to share their data with other parties.

- Deleting user location data after it is no longer necessary to perform contact tracing.

- Alignment with the Fair Information Practice Principles.

- Using open-source software to foster trust in the app's privacy protection claims.

## Are People Who Are Infected With COVID-19 Forced To Share Their Location History?

Requiring people who are infected or potentially infected to track their movements and disclose their contacts achieves the highest degree of efficacy in contact tracing within a community. However, if residents cannot choose to at least selectively withhold their information, they may be <u>stigmatized, persecuted, or exploited</u> by malicious actors on the basis of their data.  Voluntary reporting respects users' rights to privacy and to informed consent. It encourages app developers to include safeguards that reduce the risk for abuse of sensitive data. However, when individuals who become infected refuse to share their contact-tracing data, the accuracy of contact tracing declines, potentially contributing to misinformation and a false sense of security.

We believe that no one should be forced to relinquish highly sensitive personal data. We dislike solutions that require potential users to consent to share their data if they become infected in order to access information about whether or not they have crossed paths with someone who was infected.  Incentives such as those outlined in the following sections should be implemented to encourage users who become infected to share their data.

## Are Healthy People Required To Use The App?

People who are healthy should also proactively choose to use a contact-tracing app rather than being mandated to do so. Potential users should be encouraged to do so by incentives, such as the opportunity to take control of their information to benefit their health, strong privacy protection policies, trust in the app's developers, clear communication, and informed consent.

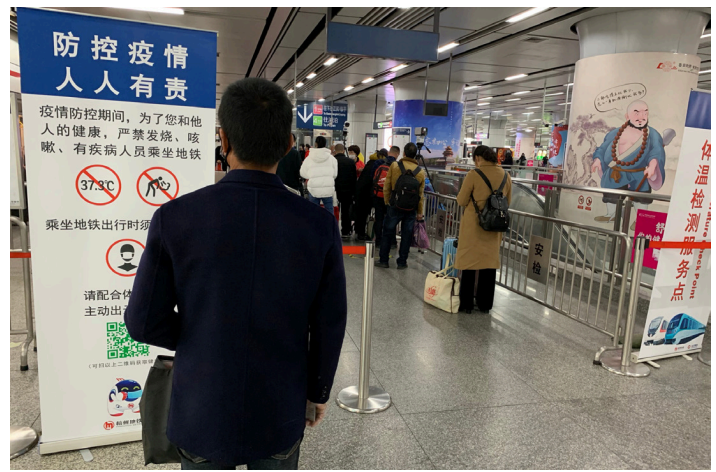

*"Alipay Health Code (China): An example of a contact-tracing approach with mandatory participation tied to freedom of movement*

*https://www.nytimes.com/2020/03/01/ business/china-coronavirus-surveillance.html*

## Who Is Supporting The App?

In order to roll out a contact-tracing app on a global scale, three groups must work together: a substantial team to create and promote the app; large, trusted institutions to support development and deployment of the app; and local, on-the-ground partners in the various communities in which the app is deployed. **Contact-tracing apps are tools, not complete solutions. Disease containment utilizing these tools requires multidisciplinary collaborations across the technology, healthcare, public health, and government sectors.** We are working hard to create these partnerships for COVID Safe Paths and look for such partnerships in other apps we evaluate.

Among those partnerships teams should be seeking to build are:

- Cloud players (AWS, Azure, GCP, etc.)

- Mobile carriers and local telecommunications providers.

- Partnerships with health authorities; these partnerships are particularly important in light of app store requirements for all apps addressing the COVID-19 pandemic to have the support of a health organization

- Government agencies

- Local public health workers and healthcare providers: contact-tracing apps will only succeed if those who crossed paths with someone who became infected can receive guidance and support from local providers on what steps to take to protect themselves and their families.

- Current contact tracers; integrating into the current contact-tracing protocol increases the effectiveness of a contact-tracing app within a community

- Non-profit organizations and academic institutions

## How Will The App Be Deployed?

We see apps aiming to deploy at a variety of levels, from a single city to an entire nation to those aiming for a global reach. Regardless of the level at which they are deployed, contact-tracing apps must be paired with existing infrastructure in order to support a successful containment strategy. Public health officials and healthcare providers must be ready to answer user questions, offer testing, or provide advice about what to do if someone has been exposed to a person with COVID-19. The resources and support necessary to follow this advice must also be made available. We look for well-considered deployment strategies with aggressive outreach to local partners. For this reason, we are building not only a contact-tracing app, but also Safe Places, a web-based tool for public health officials working to contain the COVID-19 pandemic.

It is also worth noting that as global travel resumes, cross-communication between apps operating in different regions will be necessary to achieve global containment of COVID-19. We look for teams that are thinking ahead and building the technological foundation for this collaboration into their application.

## What Steps Are Being Taken To Address Challenges In Scaling The Solution?

Taking any software tool from idea to widespread solution requires the team to think creatively. Contact-tracing apps gain value with each additional user. Many approaches to encouraging user adoption exist, and good teams will use a variety of them. A few steps we encourage are:

- Fostering trust

- Developing key partnerships, including with community officials who can help drive local support for the solution

- Creating solutions that meet the needs of public health officials responding to the pandemic

- Focusing on the needs of the users

- Providing value to the user during a contact-tracing interview even if they choose not to download the app before they have been diagnosed with COVID-19

## What Incentives Are In Place To Promote Deployment And Adoption?

Contact-tracing apps need a strong value proposition for each stakeholder—the healthy user, the person who is infected, the public health worker responsible for contact tracing, the public health authority responsible for the community's response to the pandemic, and government officials tasked with coordinating the local or national response to COVID-19. As an example, the incentives for each stakeholder from the Safe Paths solution are presented here.

### Healthy Users: The General Population

Offers an opportunity to take control and gain information. The user is able to make decisions about where they should be going and what activities are safe for their families and themselves. Users are more confident and more informed about their actual risk of spreading the disease.

### Person with COVID-19

Gives the ability to quickly and accurately share location history with public health contact tracers. Sharing their history offers an opportunity to help protect their community.

### Contact Tracers

Gives immediate relief to contact tracers. Provides a tool to more efficiently conduct interviews and gather information from patients. Increases data accuracy over current methods (e.g., remembering). Enables them to work with infected patients to quickly remove information that the patient asserts is personal, private, and/or confidential.

### Public Health Authorities

Allows more efficient and more accurate data collection and analysis about the spread of COVID-19 within their jurisdiction. Provides data to make better, more targeted recommendations for intervention to their community and to utilize limited testing resources most constructively. Offers an opportunity to communicate a personalized risk profile to each citizen, answering the question "Should I be concerned or not?" for every individual in their constituency and to closely monitor those who have the highest chance of experiencing complications from COVID-19.

### Government Officials

Faster and more accurate contact tracing allows officials to catch up with the virus and more effectively deploy resources. Rather than undifferentiated application of lock-down measures risking economic and subsequent financial collapse, officials are able to implement a differentiated approach with targeted measures as recommended by the WHO.

## What Does User Notification Of A Potential Exposure To COVID-19 Look Like?

The utmost care must be taken when notifying users of a potential exposure to COVID-19 given the serious health, economic, and social consequences of a notification. During this stressful time, clear, easy-to-understand communication reduces the possibility for the user to misjudge their situation. High-quality translations should be available for all users. Transparency about how the decision to notify the user was made helps the user and their public health officials make decisions about whether and which containment measures the user needs to undertake. Notifications should evolve to reflect advances in the understanding of disease transmission as scientists around the world continue to clarify how COVID-19 passes from person to person.

## How Does the App Prevent Fraudulent Reports?

Contact-tracing apps, particularly those that allow individuals to self-report themselves as infected, must address the risk that some people will make fraudulent reports. In some instances, a false report may be done in good faith—the person truly suspects they have COVID-19, but they have not undergone definitive testing and actually have a different virus. In other cases, bad actors may report themselves as infected with COVID-19 in order to create chaos. Storing sensitive information in an anonymized, redacted, and aggregated manner minimizes the risk of data-tampering, yet it does not eliminate the chance for human error or malicious intervention. One approach to reducing fraud requires the diagnosis to be confirmed by a healthcare provider. However, creative teams may find other ways to prevent false reports of illness.

## What Is The Policy For Correcting False Reports?

With large-scale deployment, most apps will experience an occasional false report or find an error in an otherwise correct report. Each app should develop a protocol for its response when an incorrect report is identified. Easy-to-use tools should allow all involved in reporting to quickly mark and remove errors as soon as the false report is identified. Most often, users should be notified of the change in their exposure history.

## What Happens If Identifiable Information Is Accidentally Released?

While most apps aim to obscure the identity of the person who is infected, accidental release of information sufficient to identify the person can occur on rare occasions, similar to accidental release of protected health information. These low risks should be communicated to the users during the consent process. A process for quickly removing identifiable information from public access should be in place.

## How Does The App Address The Possibility Of Panic Or Risky Behavior Among Users?

Notification of a potential exposure to COVID-19 will be frightening to many, particularly those at increased risk for serious complications, and may lead to panic among users. Large groups of people seeking medical evaluation or demanding testing could quickly overwhelm an already strained healthcare system. We have seen panic related to the pandemic lead to hoarding and vigilantism. Conversely, users who are not notified of a potential exposure may assume they are at no risk to catch COVID-19 and disregard critical social distancing and hygiene recommendations. Any contact-tracing solution will need to provide users with accurate information to reduce the chance for panic or risky behavior. When reviewing an app, we look for the following:

- Clear, easy-to-understand, culturally appropriate communication with the user

- Engagement of epidemiologists, public health officials, and healthcare providers, both as core members of the decision-making team and as local partners within the community to which the app is deployed, in order to provide assessment and recommendations to people who may have been exposed to COVID-19

- Measures to prevent individuals from falsely reporting themselves infected and thoughtful consideration of how a person reported to be infected is confirmed to have COVID-19

- Use of both GPS and Bluetooth systems, utilizing the strengths of each technology

- Creative algorithms that reduce the chance that insignificant exposures are flagged

## What Relationships Does The Team Have With Epidemiologists And Health Officials?

Contact-tracing apps should be viewed as a tool to be utilized by experts in infectious disease control. Epidemiologists, public health officials, and healthcare providers must be core members of any team designing and implementing a contact-tracing app. We look to see that such experts are included as team members, mentors, and strategic partners.

## How Does The Solution Interact With Public Health Officials?

Ideally, contact-tracing apps should fit into the current care pathway. One of the leaders in this area is TraceTogether in Singapore, which supports a contact-tracing process put in place long before the app was ready. TraceTogether uses Bluetooth to identify nearby phones with the app installed and tracks both proximity and timestamps. If a person is diagnosed with COVID-19, they can choose to allow the Ministry of Health to access their TraceTogether data, which is then used by the manual contact-tracing team to alert those who may have been exposed. The manual contact-tracing team then alerts those who may have been exposed.

We also aim to lead in this area with the development of COVID Safe Places, a web-tool allowing public health officials to work more quickly, collect better data, and better respond to what is happening in their community. We are partnering with public health workers around the world to deploy COVID Safe Places.

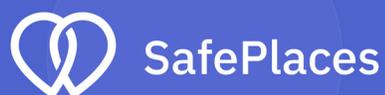

## Safe Places

A browser-based mapping tool for contact tracers to more efficiently interview infected patients and create anonymized maps and data files of public places and times where the infected patient has been.

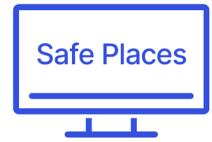

## COVID Safe Paths

COVID Safe Paths is a smartphone app that enables users to download aggregated and anonymized, infected patient location trails and simply compare them on their phone with their location history to see if they have been in close proximity to individuals who have subsequently learned that they are infected.

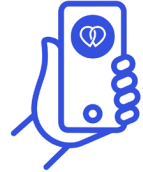

## Health Officials

When diagnosed with COVID-19, individuals can share their location history with public health officials. Together they "redact" private information, such as their home address.

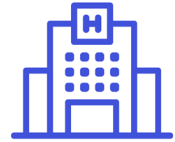

## Privacy

Safe Places creates a reliable tool and infrastructure for public health professionals, while COVID Safe Paths reduces the risk of privacy violations by replacing centralized storage of sensitive data with time-limited storage of data on the user's own device and requiring user consent for data sharing; hence avoiding a surveillance state.

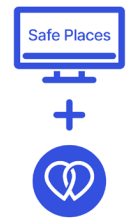

## How Will The Team Measure Its Impact On The Pandemic?

The success of any contact-tracing program should be measured in lives saved. Lives are saved both by a reduction in the spread of disease and by a reduction in the psychosocial and economic consequences of widespread quarantine actions. Quantitative analysis of the effect of this new technology should be undertaken—not only to allow for further improvements during the current COVID-19 pandemic, but also to better address the next outbreak of infectious disease.   In addition to collecting real-world data about the impact of contact-tracing apps, teams should work to communicate their success to the public. If the apps are effective in helping to control the pandemic, the public may <u>fail to notice</u> the extent to which their use was critical to the community's ability to control the spread of disease.

# Conclusion

The COVID-19 pandemic will not last forever. If we falter in our response and choose digital contact-tracing tools that compromise individual privacy for efficacy, the consequences will extend long after the last store has reopened and the last child has returned to school.  We believe privacy does not have to be compromised in order to reduce new infections and slow the spread of disease.  We are building COVID Safe Paths with privacy protection at the forefront for this pandemic and the next.   Here, we have begun to detail the key questions that should be asked as we evaluate contact-tracing apps developed and deployed against the COVID-19 pandemic.  We plan to continue this discussion and are committed to serving as a resource for countries, states, cities, and individuals throughout the world.

We welcome additions to and modifications of this report and analysis. To submit a change please email info@pathcheck.org

## Authors
Ramesh Raskar | Greg Nadeau | John Werner | Rachel Barbar | Ashley Mehra | Gabriel Harp | Markus Leopoldseder | Bryan Wilson | Derrick Flakoll | Praneeth Vepakomma | Deepti Pahwa | Robson Beaudry | Emelin Flores | Maciej Popielarz | Akanksha Bhatia | Andrea Nuzzo | Matt Gee | Jay Summet | Rajeev Surati | Bikram Khastgir | Francesco Maria Benedetti | Kristen Vilcans | Sienna Leis | Khahlil Louisy

**Design by:** Tony Pham

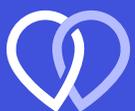 COVID SafePaths